# A transfer learning enhanced the physics-informed neural network model for vortex-induced vibration

Hesheng Tang [a, *], Hu Yang [a], Yangyang Liao [a], Liyu Xie [a,]

a. *Department of Disaster Mitigation for Structures, College of Civil Engineering, Tongji University, Shanghai, 200092, China*

**Abstract:** Vortex‑induced vibration (VIV) is a typical nonlinear fluid-structure interaction phenomenon, which widely exists in practical engineering (the flexible riser, the bridge and the aircraft wing, etc). The conventional finite element model (FEM)-based and data-driven approaches for VIV analysis often suffer from the challenges of the computational cost and acquisition of datasets. This paper proposed a transfer learning enhanced the physics-informed neural network (PINN) model to study the VIV (2D). The physics-informed neural network, when used in conjunction with the transfer learning method, enhances learning efficiency and keeps predictability in the target task by common characteristics knowledge from the source model without requiring a huge quantity of datasets. The datasets obtained from VIV experiment are divided evenly two parts (source domain and target domain), to evaluate the performance of the model. The results show that the proposed method match closely with the results available in the literature using conventional PINN algorithms even though the quantity of datasets acquired in training model gradually becomes smaller. The application of the model can break the limitation of monitoring equipment and methods in the practical projects, and promote the in-depth study of VIV.

## 1. Introduction

Vortex-induced vibration (VIV) is a typical fluid-structure interaction phenomenon, which can be seen universally in practical engineering, such as resonance of cable structures in the wind field, sustained vibration of riser in the deep sea (Williamson and Govardhan, 2008; Liu et al., 2021; Matin et al., 2018). When the fluid passes over the surface of structure, two sides of the structure will occur vortex detachment alternately, which in turn generates periodic fluid forces on the structure. Under the action of fluid forces, the structure starts to vibrate periodically and strongly, resulting that structure suffers different degrees of fatigue damage (Wang et al., 2020). Therefore, various scholars use kinds of methods to research related problems, and apply them into project cases. Nowadays analyzing VIV with computational fluid dynamics (CFD) has made great progress (Martini et al., 2021). However, it is difficult to break through the limitation of computation capacity and moving mesh precision (Liu et al., 2021). Thereby, it is essential and urgent to improve the computational efficiency of VIV.

With the development of artificial intelligence (AI), the deep learning has gradually been applied to research VIV problems (Kim et al., 2021). Lim and Kim (2021) achieved the detection of fatigue damage of propeller in VIV based on the data-driven neural network. Wong and Kim (2018) proposed a more simplified neural network to forecast VIV fatigue damage of top tensioned riser (TTR). Wu et al. (2018) used the neural network to establish surrogate model of VIVACE converter in VIV region. In order to reduce the quantity of datasets required for training model and improve the robustness of the neural network, the PINN was used to study VIV in place of the data-driven neural network (Jin et al., 2021). Raissi et al. (2019) solved forward and inverse problems by embedding partial differential equations (PDE) into the conventional deep neural network instead of discrete numerical computation. Cheng et al. (2021) employed PINN based on Reynolds Average Navier-Stokes equations to analyze VIV and wake-induced vibration (WIV) of the cylinder, and verified the high efficiency of PINN in solving similar problems. Bai and Zhang (2021) added the viscosity coefficient into physics equations, extending the study of VIV from laminar flow to turbulent flow using PNS-PINN. Raynaud et al. (2021) proposed extended framework of PINN (ModalPINN) by means of encoding the approximation of Fourier mode shapes, and improved the generalization ability of surrogate model with limited data. Sun et al. (2020) established a framework of the deep learning, which includes physics equations, initial conditions and boundary conditions, to forecast the information of the flow field without simulation. Nevertheless, it is full of challenges to monitor data of VIV in actual projects (Dan and Li, 2021). It requires finding another way to improve such a bad situation.

In recent years, as a brand-new machine learning approach, the transfer learning is presented to increase the computational efficiency of the deep learning, which provides a breakthrough idea for further study of VIV (Zhuang et al., 2021). The essence of transfer learning is to extract common characteristics knowledge from the source domain to complete new learning tasks in the target domain so as to reduce the cost of datasets and time (Behnam et al., 2020). There are mainly two categories of transfer learning: the isomorphic transfer learning and the heterogeneous transfer learning (Pan and Yang, 2009). The isomorphic transfer learning includes dataset shift, domain adaptation and multi-tasks learning (Shu and Li, 2021). The heterogeneous transfer learning includes cross-modal learning and cross-category learning (Rahman and Islam, 2021). Zafer and Wang (2020) proposed a method based on transfer learning, which can make the distribution of statistical parameters of random processes close to each other in different time period, and achieve the forecast of time-dependent reliability. Chen et al. (2020) and Kaur et al. (2020) combined transfer learning with deep convolutional neural network (DCNN), to complete image identification and classification respectively. Wu et al. (2020) monitored bearing failure accurately by using adaptive deep transfer learning based on features. Gupta et al. (2021) put forward a cross property framework of the deep transfer learning, to enhance the accuracy of forecast on small materials data. Ye and Dai (2021) achieved time series prediction with a small quantity of datasets using transfer learning, and solved the problem of missing data.

In this paper, to acquire high-precision forecast results with lower cost of datasets and time, the transfer learning was employed to enhance the conventional PINN model for VIV analysis. By means of pre-training the labeled datasets in the source domain and fine-tuning the architecture of the deep neural network, the transfer learning is able to complete training objectives with a small quantity of the labeled datasets in the target domain. In the section 1, the research status of VIV and transfer learning is elaborated in detail. In addition, the physic model, the PINN model and the transfer learning model for VIV are introduced in the section 2 to section 4 respectively. To verify the performance of the proposed model, the experimental datasets are utilized to compute four different working conditions in the numerical examples in the section 5. The accuracy of forecast results and the cost of datasets and time are compared respectively, as it turns out, the whole learning tasks are completed with lower cost of datasets and time in the target domain by transferring common characteristics knowledge of VIV. All the working conditions accurately forecast the velocity and pressure of the flow field, as well as the vibration displacement of the structure and the fluid force coefficient. It is proved that the transfer learning has a significant enhancement to the conventional PINN model for VIV.

## 2. The physical model of VIV (2D)

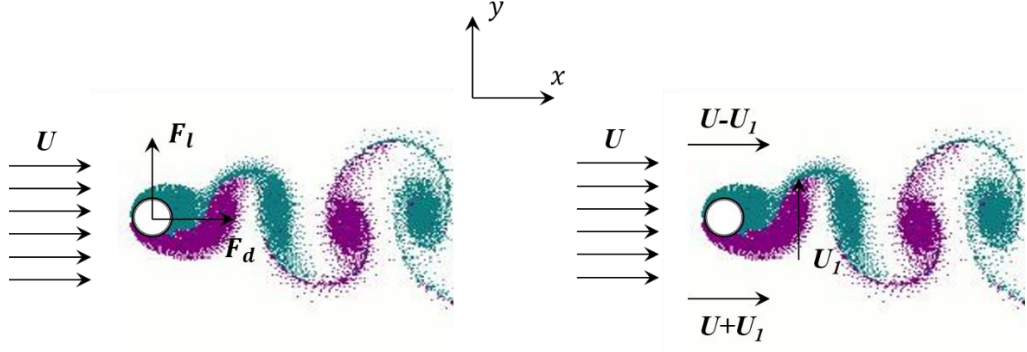

Fig. 1. The fluid forces of VIV.

As the structure occurs vortex-induced vibration in the flow field, it is subjected to the coupling forces in two different directions. The fluid lift force $F_l$ is on the cross-flow (CF) direction (*i.e.*, $y$), and the fluid drag force $F_d$ is on the in-line (IL) direction (*i.e.*, $x$) (as show in Fig. 1). The initial velocity is denoted by $U$. The fluid circulation is produced by the rotation of the fluid around the structure, which is represented by $U_1$. The lift force $F_l$ is generated when the velocity distribution on the upper and lower surfaces of the structure is no longer symmetrical due to the effect of fluid circulation. And the viscidity of natural fluid generates the drag force $F_d$.

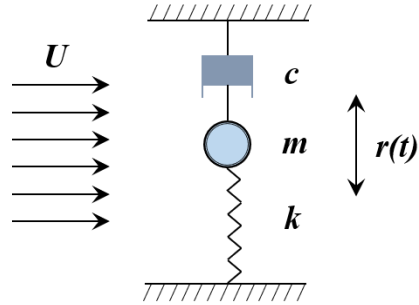

Fig. 2. The simplified physical model of the unidirectional VIV.

Currently, most studies solely focus on the unidirectional vibration (CF direction), because the response amplitude of IL direction is extremely minimal compared to that of CF direction. As a result, to simplify the physics equation embedded to the deep neural network, the VIV physical model is converted into a unidirectional mass-spring-damp elastic system, only monitoring vibration displacement of the structure on the CF direction. The simplified physical model of VIV is illustrated in Fig. 2. And the physics equation of the structure on the CF direction can be expressed as Eq. (1):

$$m\frac{d^2r}{dt^2}+c\frac{dr}{dt}+kr=F_l \tag{1}$$

Where *m*, *k* and *c* are the mass, stiffness and damping of the structure respectively, *t* represents time and *r(t)* represents vibration displacement of the structure on the CF direction.

For the flow fields at small Reynold number, the fluid can be regarded incompressible, which implies the fluid's density can remain constant at all times. Therefore, the properties of the fluid should fulfill the following assumptions:

(1) The fluid conforms to the constitutive equation of the ideal Newtonian fluid;

(2) The flow field is always in the laminar flow state;

(3) The boundary conditions of flow field are based on the assumption of non-slipping.

The research only focuses on the correlation computation of two-dimensional (2D) VIV. And the incompressible Navier-Stokes equations, such as the continuity differential equation and the motion differential equation, are used to explain 2D fluid motion. The equations are given explicitly by Eq. (2)-Eq. (4).

The continuity differential equation can be expressed as:

$$\frac{\partial u}{\partial x}+\frac{\partial v}{\partial y}=0 \tag{2}$$

The motion differential equation can be expressed as:

$$\frac{\partial u}{\partial t}+u\frac{\partial u}{\partial x}+v\frac{\partial u}{\partial y}+\frac{\partial p}{\partial x}-\frac{1}{Re}(\frac{\partial^2 u}{\partial x^2}+\frac{\partial^2 u}{\partial y^2})=0 \tag{3}$$

$$\frac{\partial v}{\partial t}+u\frac{\partial v}{\partial x}+v\frac{\partial v}{\partial y}+\frac{\partial p}{\partial y}+\frac{\partial^2 r}{\partial t^2}-\frac{1}{Re}(\frac{\partial^2 v}{\partial x^2}+\frac{\partial^2 v}{\partial y^2})=0 \tag{4}$$

Where (*x, y*) represents the coordinates of the flow filed, *u* (*t, x, y*) and *v* (*t, x, y*) represent the velocity in IL and CF directions of the flow field respectively, *p* (*t, x, y*) is the pressure of the flow filed. The initial fluid velocity *U*, the fluid density $\rho$, characteristic length of the structure *d* and coefficient of the kinetic viscosity $\mu$ all contribute to the dimensionless number *Re*, which represents the state of fluid flow.

In the process of VIV, there is significant interaction between the flow field and structure, which is caused by the combined action of the lift force $F_l$ and the drag force $F_d$. To be more specific, the boundary conditions (BC) and initial conditions (IC) of the fluid-structure interaction surface, as well as the distribution of fluid velocity and pressure, influence the coupling forces acting on the structure.

Based on the velocity and pressure (*u*, *v*, *p*) of points on the fluid-structure interaction surface, the interaction forces can be acquired. Eq. (5)- Eq. (6) show how the interaction forces are calculated using an integrated method based on the BC/IC of the interaction surface. It is obvious that the combination of Eq. (1) and Eq. (5) depicts the significant coupling of VIV on the CF direction.

$$F_l = \oint [-pn_y + 2\text{Re}^{-1} v_y n_y + \text{Re}^{-1}(u_y + v_x)n_x] ds \tag{5}$$

$$F_d = \oint [-pn_x + 2\text{Re}^{-1} u_x n_x + \text{Re}^{-1}(u_y + v_x)n_y] ds \tag{6}$$

Where ($n_x$, $n_y$) represents the normal vector to the fluid-structure interaction surface, *ds* represents the integral arc length.

## 3. The Physics-informed neural network for VIV

### 3.1 The architecture of deep neural network

The deep neural network (DNN) is a branch of machine learning. Its architecture is composed of multiple nonlinear single-layer networks, and each layer of network is connected by a large number of neurons (Liu et al., 2017). Based on the kind of connections between neurons, the DNN can be classified as fully-connected neural networks (FCNN), convolutional neural networks (CNN) and recurrent neural networks (RNN), etc. (Deng et al., 2013). Before training of the DNN, the loss function should be constructed based on the output layer parameters.

The loss functions commonly used in the DNN include the mean squared error (MSE) and mean absolute error (MAE) (Zhang et al., 2019). There are two vital parameters in the DNN: the weight matrix $W_{m,n}$ and bias vector $b_m$. Where *m* represents the number of hidden layers, *n* represents the number of neurons in each hidden layer (Bau et al., 2020). In the course of training datasets (the data obtained by the numerical simulation, experimental and other means), $W_{m,n}$ and $b_m$ of the DNN are constantly modified to reduce the loss function to the minimized threshold value, to achieve the assigned training objectives.

The DNN adopted in this paper is fully-connected neural network (FCNN), which contains input layer, output layer and hidden layer (the hidden layer is 8 layers with 20 neurons in each layer). As is illustrated in Fig. 3, the variables of input layer are *t*, *x* and *y* respectively, and the variables of output layer are *u* (*t*, *x*, *y*), *v* (*t*, *x*, *y*), *p* (*t*, *x*, *y*) and *r*(*t*) respectively.

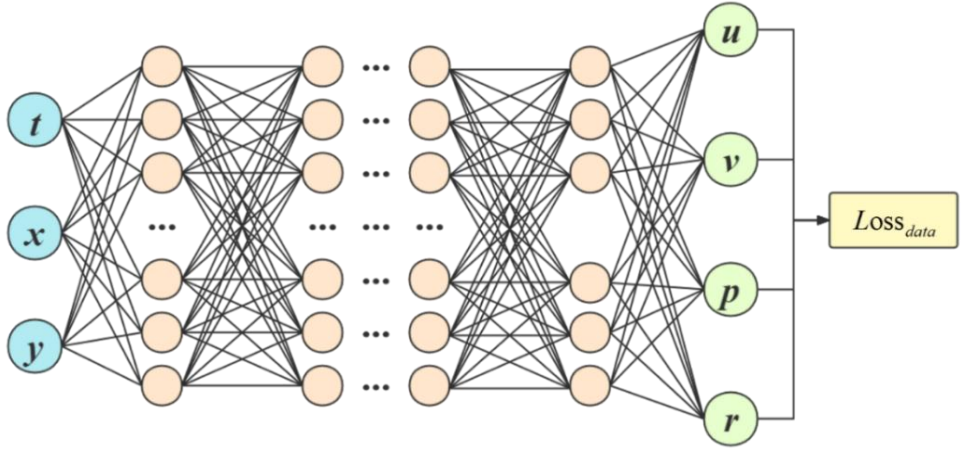

Fig. 3. The deep neural network architecture for VIV.

In the progress of signal forward propagation, $\theta(x)=\sin x$ is chosen as for the activation function, to normalize the training sample in the range 0-1 and achieve nonlinear mapping of the DNN. Meanwhile, the Adam optimization algorithm is used to achieve the target of gradient descent. The algorithm can update parameters of the DNN high-efficiently and optimize loss function of the neural network to the minimized threshold value during signal back propagation.

The data-driven loss function is expressed as $Loss_{data}$, which can be expressed as:

$$Loss_{data} = \sum_{n=1}^{N} (|u(t^n,x^n,y^n)-u^n|^2 + |v(t^n,x^n,y^n)-v^n|^2 + |p(t^n,x^n,y^n)-p^n|^2 + |r(t^n)-r^n|^2) \quad (7)$$

Where $N$ represents the total number of training datasets, $u^n$, $v^n$, $p^n$ and $r^n$ represent the real value of fluid velocity in IL direction, fluid velocity in CF direction, pressure and displacement respectively, $u(t^n, x^n, y^n)$, $v(t^n, x^n, y^n)$, $p(t^n, x^n, y^n)$ and $r(t^n)$ represent the forecast value of fluid velocity in IL direction, fluid velocity in CF direction, pressure and vibration displacement respectively.

### 3.2 The physics-informed neural network

The physics-informed neural network (PINN) is proposed recently, which embeds objective physics equations into the DNN. By means of PINN, the correct answer of nonlinear PDE can be quickly approached, to solve forward and inverse problems instead of the traditional discrete numerical methods. Compared with the traditional data-driven DNN, the PINN uses training datasets and physics information as common constraints to train the DNN (Cai et al., 2021). The automatic differentiation (AD) is applied to the computation of physics equations, to construct extra loss function to indicate the approximation degree of related physics equations, which is expressed as $Loss_{phy}$. According to recent study, the PINN can reduce the cost of datasets and time while enhancing the accuracy of eventual results, and considerably improve generalization capacity of the DNN. In this paper, the incompressible Navier-Stokes equation (2D), BC/IC of the fluid-structure interaction surface and dynamic equation of the structure (the interaction conditions) are embedded into the DNN, to solve VIV problems, which is illustrated in Fig. 4.

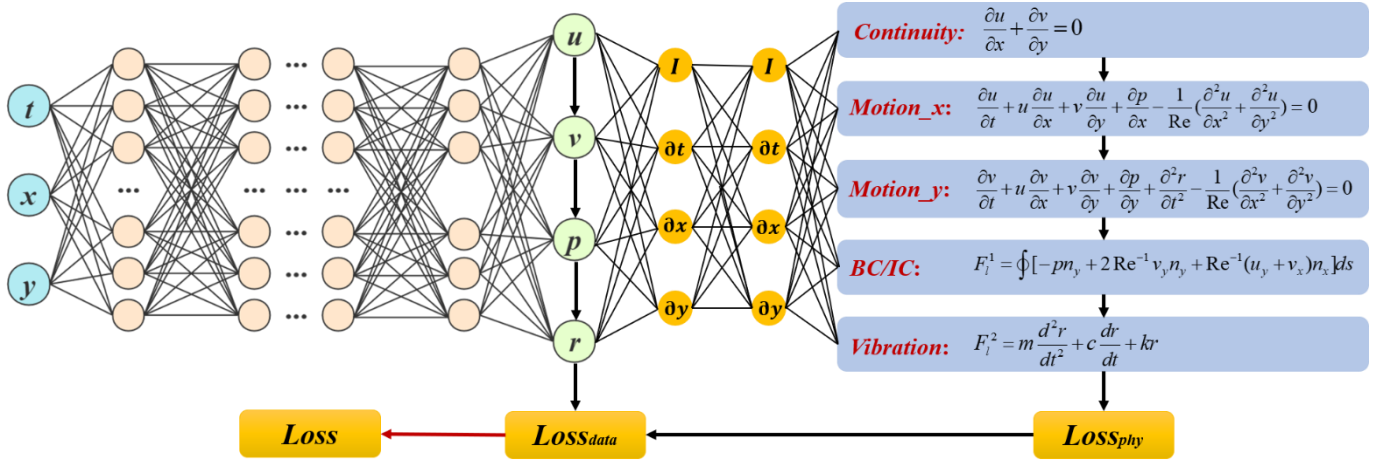

Fig. 4. The physics informed neural network for VIV.

To indicate $Loss_{phy}$ conveniently, $a_1$, $a_2$, $a_3$ are used to substitute Eq. (2)- Eq. (4), which is expressed by Eq. (8)-Eq. (10). Therefore, $Loss_{phy}$ is explicitly given by Eq. (11) to make sure the fitting of physics equations.

$$a_1 = \frac{\partial u}{\partial x} + \frac{\partial v}{\partial y} \tag{8}$$

$$a_2 = \frac{\partial u}{\partial t} + u\frac{\partial u}{\partial x} + v\frac{\partial u}{\partial y} + \frac{\partial p}{\partial x} - \frac{1}{Re}(\frac{\partial^2 u}{\partial x^2} + \frac{\partial^2 u}{\partial y^2}) \tag{9}$$

$$a_3 = \frac{\partial v}{\partial t} + u\frac{\partial v}{\partial x} + v\frac{\partial v}{\partial y} + \frac{\partial p}{\partial y} + \frac{\partial^2 r}{\partial t^2} - \frac{1}{Re}(\frac{\partial^2 v}{\partial x^2} + \frac{\partial^2 v}{\partial y^2}) \tag{10}$$

$$Loss_{phy} = \sum_{i=1}^{3}\sum_{n=1}^{N}(|a_i(t^n, x^n, y^n)|^2) + \sum_{n=1}^{N}|F_l^1(t^n) - F_l^2(t^n)|^2 \tag{11}$$

The loss function of PINN used in this paper is defined as $Loss$, which is composed of two parts: $Loss_{data}$ and $Loss_{phy}$. Thereby $Loss$ can be expressed as Eq. (12).

$$\begin{aligned} Loss = Loss_{data} + Loss_{phy} = &\sum_{n=1}^{N} \begin{array}{l}(|u(t^n,x^n,y^n)-u^n|^2 + |v(t^n,x^n,y^n)-v^n|^2 \\ + |p(t^n,x^n,y^n)-p^n|^2 + |r(t^n)-r^n|^2)\end{array} \\ & + \sum_{i=1}^{3}\sum_{n=1}^{N}(|a_i(t^n,x^n,y^n)|^2) + \sum_{n=1}^{N}|F_l^1(t^n)-F_l^2(t^n)|^2 \end{aligned} \quad (12)$$

Where $N$ represents the total number of training samples, $a_i(t^n, x^n, y^n)$ represents the computational results of Eq. (8)-Eq. (10) with the forecast values ($u$, $v$, $p$, $r$) as the dependent variables, $F_l^1(t^n)$ represents the computation results of Eq. (5) with the forecast values ($u$, $v$, $p$) as the dependent variables, and $F_l^2(t^n)$ represents the computation results of Eq. (1) with the forecast values $r(t)$ as the dependent variables. $F_l^1(t^n)$-$F_l^2(t^n)$ is to verify whether the interaction condition of the fluid motion and the structure vibration is satisfied on the CF direction.

## 4. The transfer learning enhanced the PINN model for VIV

### 4.1 The transfer learning

For the traditional machine learning (ML), well-performing supervised leaning depends on large numbers of labeled datasets and longstanding training. However, by searching for the common potential characteristics between two different domains, the transfer learning can transfer the common characteristic knowledge from domain A to domain B, to finish related but different tasks (Bengio, 2012). And the domain A is termed as the source domain, the domain B is termed as the target domain. By training datasets in the source domain, the related parameters representing the characteristics knowledge of the DNN can be extracted. Therefore, the datasets in the source domain can be taken as training basis of datasets in the target domain. In this way, the transfer learning can not only effectively reduce the number of datasets and computation time required for training the DNN, but also ensure the same (even better) forecast accuracy as the traditional supervised learning.

The differences between the traditional supervised learning and the transfer learning when performing related but different tasks are demonstrated in Fig. (5). As is illustrated in Fig. 5(a), for different learning tasks, the traditional supervised learning is bound to remark the whole datasets before each training session and extract distinct characteristics knowledge from the DNN, which means that every task has to start from scratch. Nevertheless, As is illustrated in Fig. 5(b), with the help of the transfer learning, the common characteristic knowledge can be extracted from the training model of datasets in the source domain. By sharing the common characteristics knowledge between the source domain and the target domain, it is needless to remark datasets in the target domain overmuch, and spend much time training model, which saves the cost of datasets and time vastly.

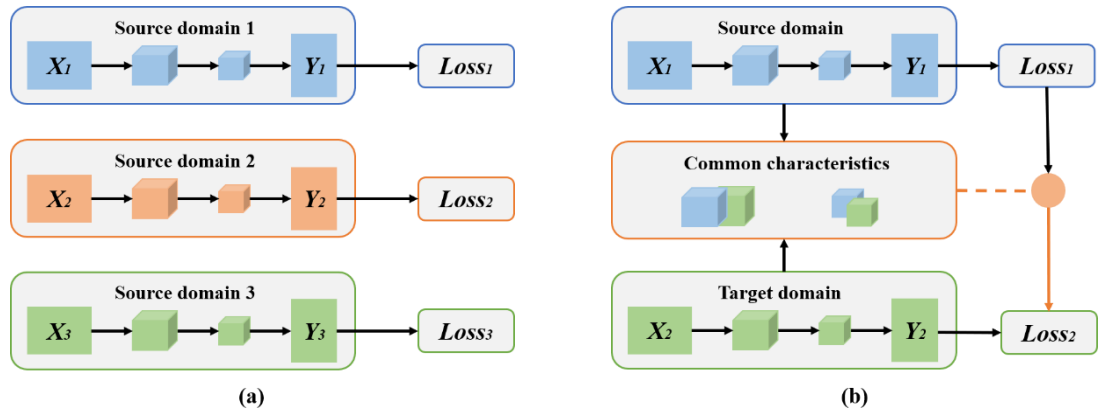

Fig. 5. The differences in completing different learning tasks. (a) the traditional supervised learning; (b) the transfer learning.

### 4.2 The transfer learning with the PINN

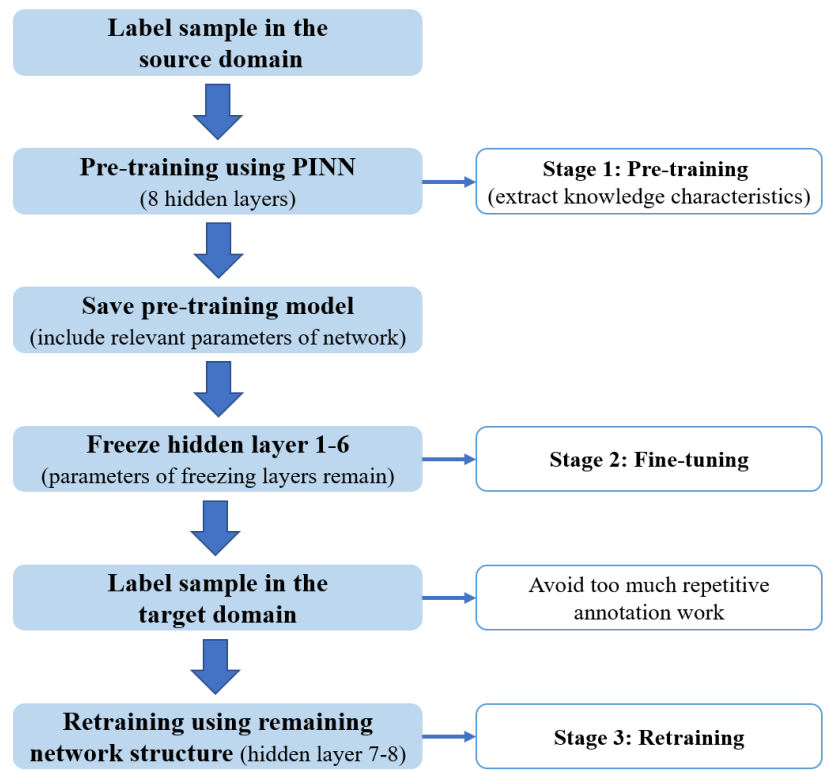

Fig. 6. The operation procedure of the transfer learning.

In the practical engineering, monitoring the velocity and pressure of the flow field with restricted monitoring equipment and methods is exceedingly challenging.

Even though solving VIV problems by PINN, annotating datasets and training the DNN aren't a small amount of work. As a result, in order to improve the applicability of PINN, it is essential and vital to decrease the data cost, as well as the time cost.

In order to further optimize the computational process of the PINN, the transfer learning is combined with the DNN for solving the problems of VIV (Goswami et al., 2020). In this paper, the isomorphic transfer learning with domain adaptation (only consider transfer learning in the same flow field) is adopted to solve the VIV problems, which includes three pivotal transaction stages: pre-training, fine-tuning and retraining, as is illustrated explicitly in Fig. 6.

Before starting the transfer learning, the datasets of the source domain and target domain should be defined according to training objectives. The training objectives for VIV analysis are to reduce the cost of datasets and time, obtain the high-precision forecast results and improve the computational efficiency of the PINN to some extent.

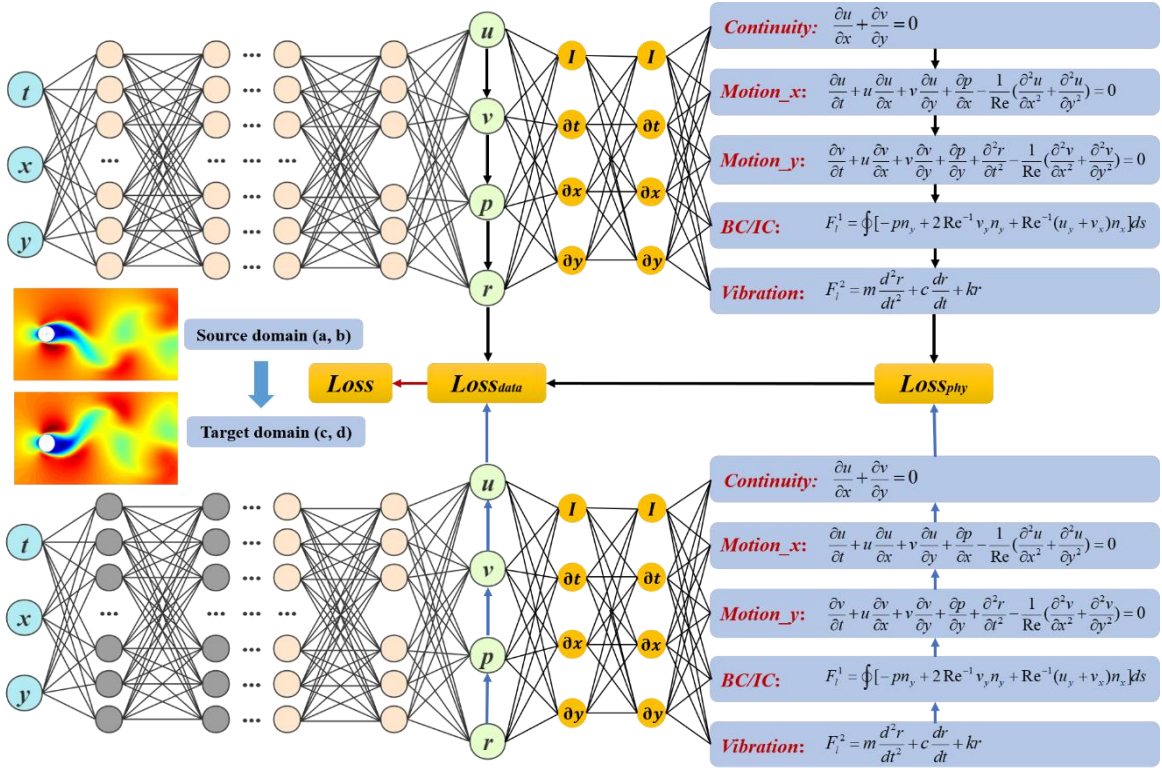

Fig. 7. Fine-tuning for PINN in the transfer leaning (Gray neurons represents the frozen neurons, and frozen layers are the hidden layer 1-6) .

Firstly, the whole network architecture of the PINN is used to pretrain datasets in the source domain, to gradually extract the characteristic parameters of the fluid motion and the structure vibration. Then the network architecture is fine-tuned based on preserving related parameters of the pre-trained DNN. As is shown in Fig. 7, the hidden layer 1-6 are totally frozen, which means parameters of these frozen layers will remain unchanged in the process of retraining. At last, the rest of network architecture (input layer, output layer and hidden layer 7-8) is used to retrain datasets in the target domain on a small scale, to achieve the transfer training with lower cost of datasets and time. It is essential and vital to promote such a frugal approach to solve VIV problems.

## 5. Numerical examples

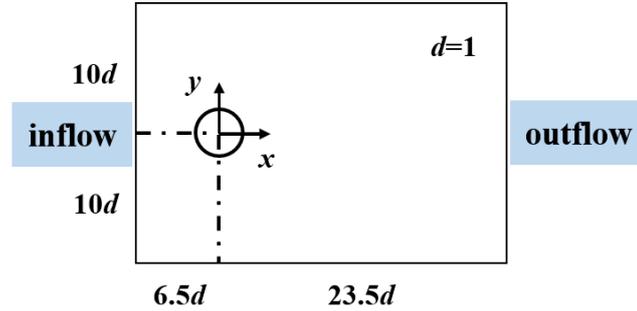

Fig. 8. The flow field in the numerical examples.

The experimental datasets used in the numerical examples are come from Raissi el al. (2019), which are measured by the particle image velocimetry (PIV). The PIV is a hydromechanical velocity measurement method developed at the end of 1970s, which is transient, multi-point and non-contact (Li et al., 2020). As show in Fig. 8, the flow field monitoring area is [-6.5$d$, 23.5$d$] * [-10$d$, 10$d$] in the 2D rectangular coordinates, and the structure is in the center of the coordinate. In the figure, $d$ represents the characteristic length of the structure and its value is 1. When it comes to the BC of the flow field, $x/d$=-6.5 is defined as the inflow boundary, $x/d$=23.5 is defined as the outflow boundary, the remaining boundaries are set as the symmetric non-slipping boundaries, and the boundary pressure is zero. The numerical examples are to study VIV in the laminar flow with value $Re$=100, so the viscous coefficient isn't embedded into the DNN.

In the numerical examples, the total monitoring time is 14 s and the time step is 0.05 s, which means monitoring equipment is going to collect 280 moments of training datasets. In each time step, the equipment monitors the velocity and pressure ($u$, $v$, $p$) of 14400 points in the flow field, and the vibration displacement $r$ of the structure as well. Before starting the transfer learning, the whole datasets are divided into two parts evenly. As is illustrated in Fig. 9, datasets from 0 to 7s are defined as the source domain to complete the pre-training task, and datasets from 7 to 14s are defined as the target domain to complete the retraining task.

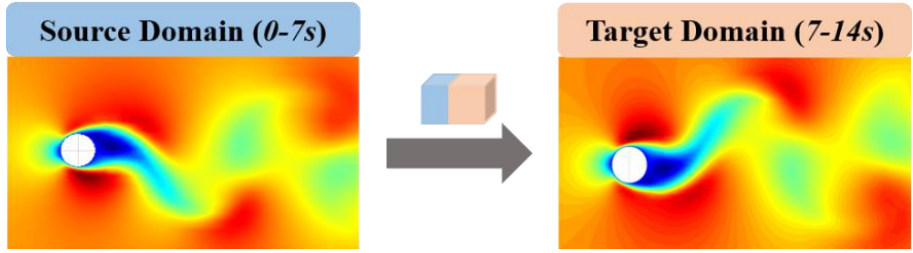

Fig. 9. The definition of the source domain and target domain.

To test the enhancement effect of the transfer learning model for VIV, the following four working conditions are computed respectively: (1) training all the datasets in the target domain by the PINN; (2) training 1/2 datasets in the target domain by the transfer learning; (3) training 1/4 datasets in the target domain by the transfer learning; (4) training 1/8 datasets in the target domain by the transfer learning. All the working conditions are shown in Table 1 explicitly.

Table 1. Introduction of four working conditions.

| Serial number | Number of monitoring points per time step | Total time steps of target domain | Training method |
|---|---|---|---|
| 1 | 14400 | 140 | PINN |
| 2 | 7200 | 140 | Transfer learning |
| 3 | 3600 | 140 | Transfer learning |
| 4 | 1800 | 140 | Transfer learning |

Table 2. The setting of hyperparameters in training neural network.

| Phase of training | The number of epochs | Batch size | Learning rate |
|---|---|---|---|
| Stage 1 | 200 | 10000 | 1.00E-03 |
| Stage 2 | 300 | 10000 | 1.00E-04 |
| Stage 3 | 200 | 10000 | 1.00E-05 |

The setting the hyperparameters of the deep learning reasonably is the key to obtain high-precision forecast results, the following hyperparameters are especially important: the epoch, batch size and learning rate. Epoch represents the total number of training iterations. Batch size affects the accuracy of determining the optimal gradient descent direction and the speed of network convergence. Learning rate is also called step size, which determines the speed of updating parameters of the DNN. To compare the eventual results more clearly, the four working conditions in this paper all adopt the same hyperparameters setting, which is shown in Table 2.

According to the eventual results, the four working conditions all forecast motion of the flow field and vibration displacement of the structure with high accuracy in the target domain. Then the forecast results of four working conditions are compared, as are shown in Fig. 10 and Fig. 11, the velocity and pressure (*u*, *v*, *p*) of the flow field at the 8th and 11th second are displayed respectively. Meanwhile, the vibration displacement time history of the structure is shown in Fig. 12. In order to demonstrate time-saving ability of the transfer learning, Table 3 shows how long it takes to compute four working conditions respectively.

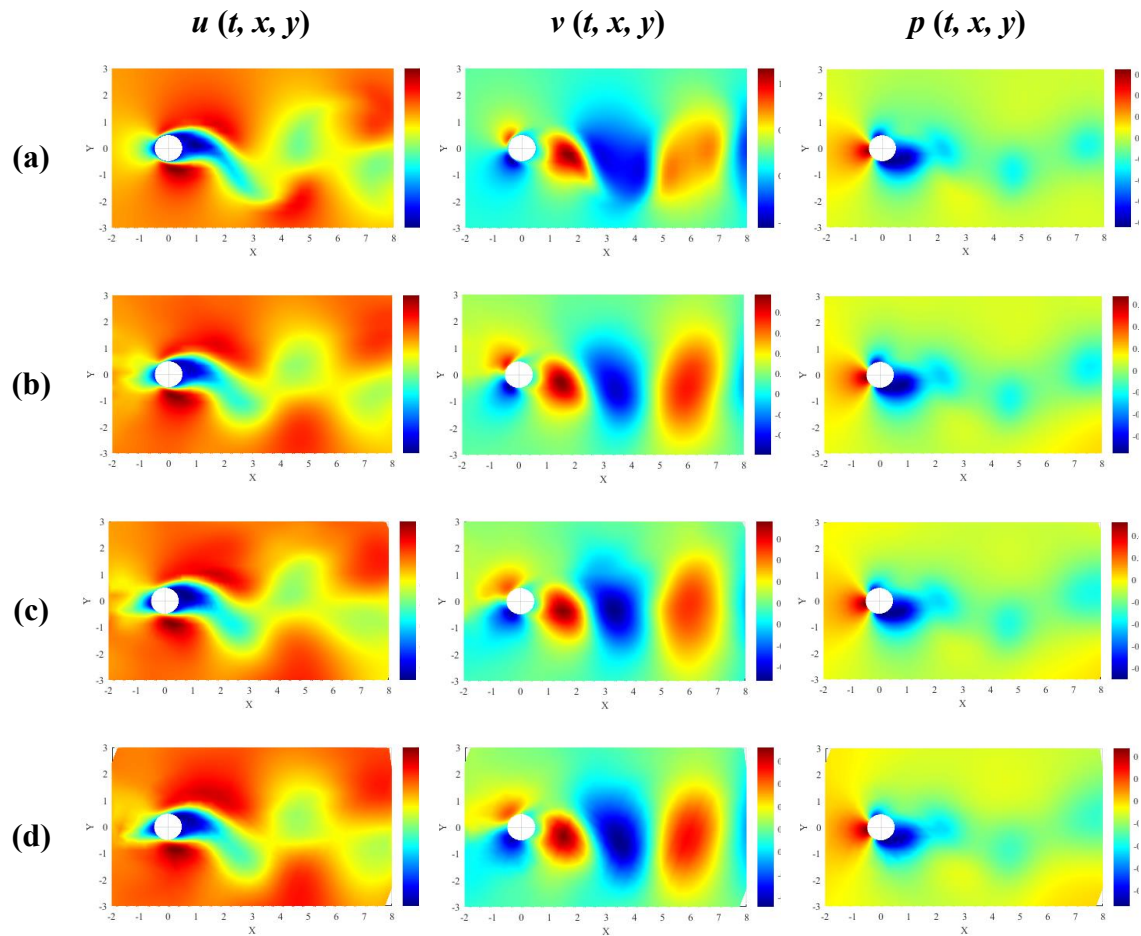

Fig. 10. The motion of the flow field at the 8th second. (a) condition 1; (b) condition 2; (c) condition 3; (d) condition 4.

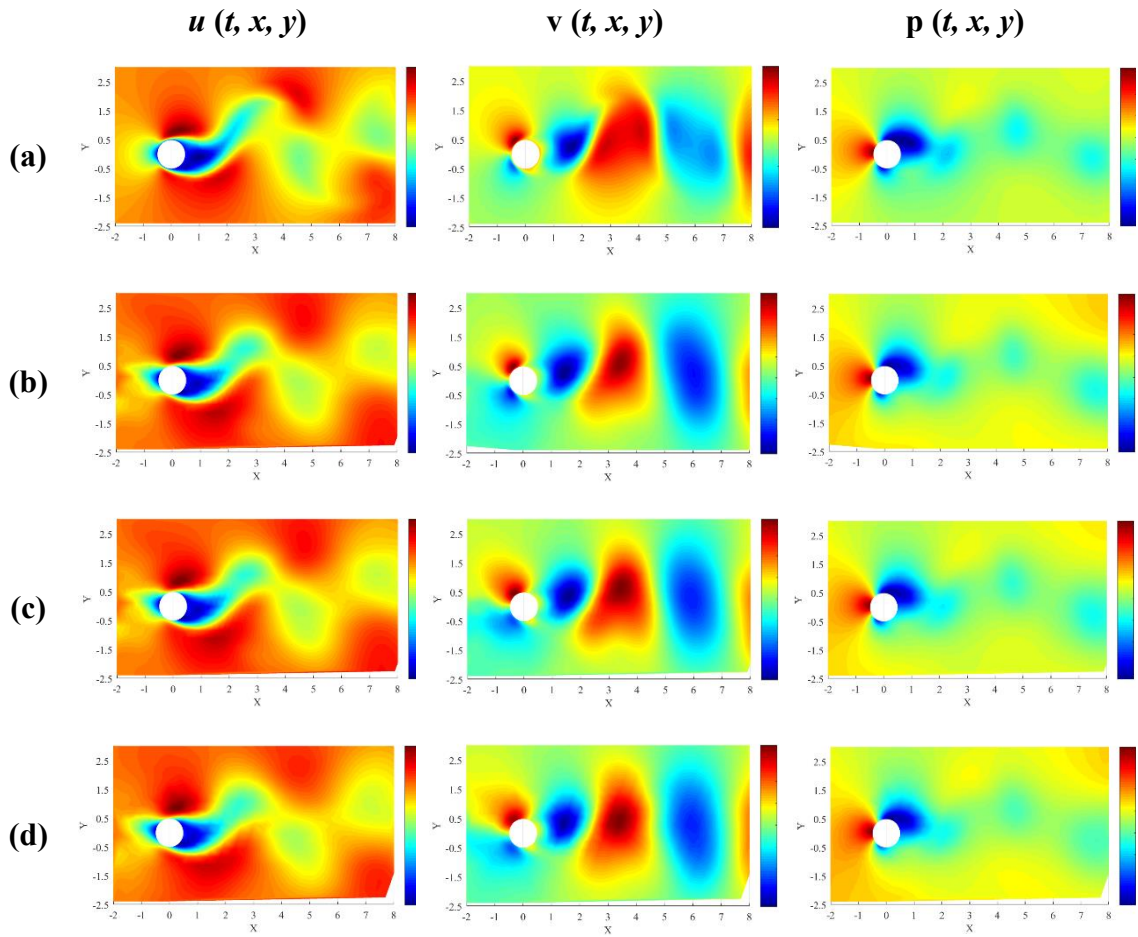

Fig. 11. The motion of the flow field at the 11th second. (a) condition 1; (b) condition 2; (c) condition 3; (d) condition 4.

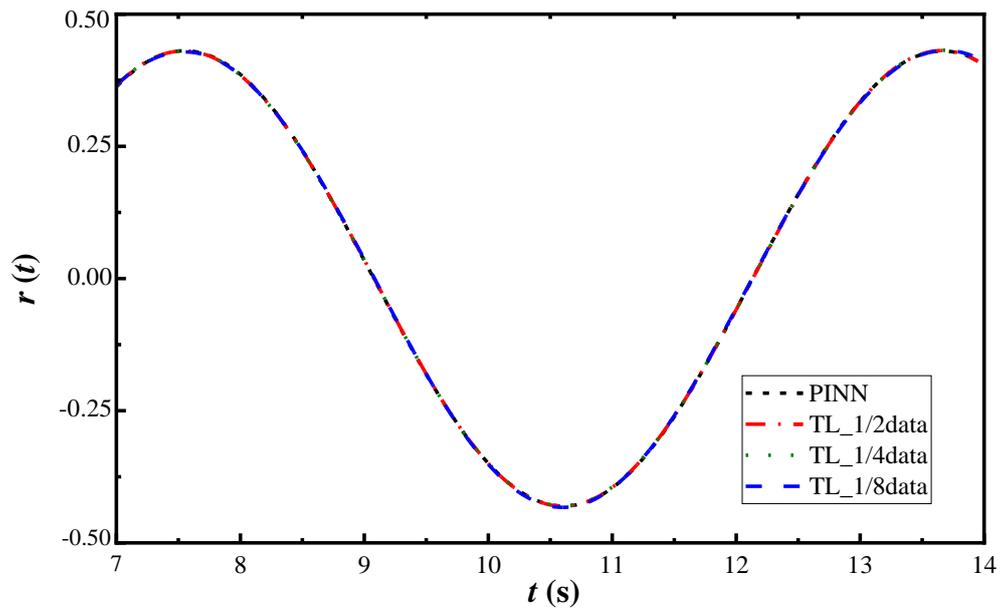

Fig. 12. The curve of vibration displacement and time for the structure.

Table 3. The cost of time for different working conditions

| Serial number | Time cost (h) |
|---|---|
| 1 | 48.7 |
| 2 | 22.9 |
| 3 | 10.3 |
| 4 | 5.4 |

To compare the accuracy of forecast results, the root mean squared error (RMSE) is used to measure the error between the forecast value and the true value, which is expressed as Eq. (13).

$$RMSE = \sqrt{\frac{\sum_{i=1}^{N}(X_{fore,i} - X_{true,i})^2}{N}} \tag{13}$$

Where $N$ presents the number of monitoring points, $X_{fore,i}$ represents the forecast value, and $X_{true,i}$ represents the true value. The smaller RMSE is, the higher accuracy of forecast results. The RMSE for the forecast results of IL velocity $u$, CF velocity $v$, pressure $p$ and vibration displacement $r$ are shown clearly in Fig. 13 and Fig. 14.

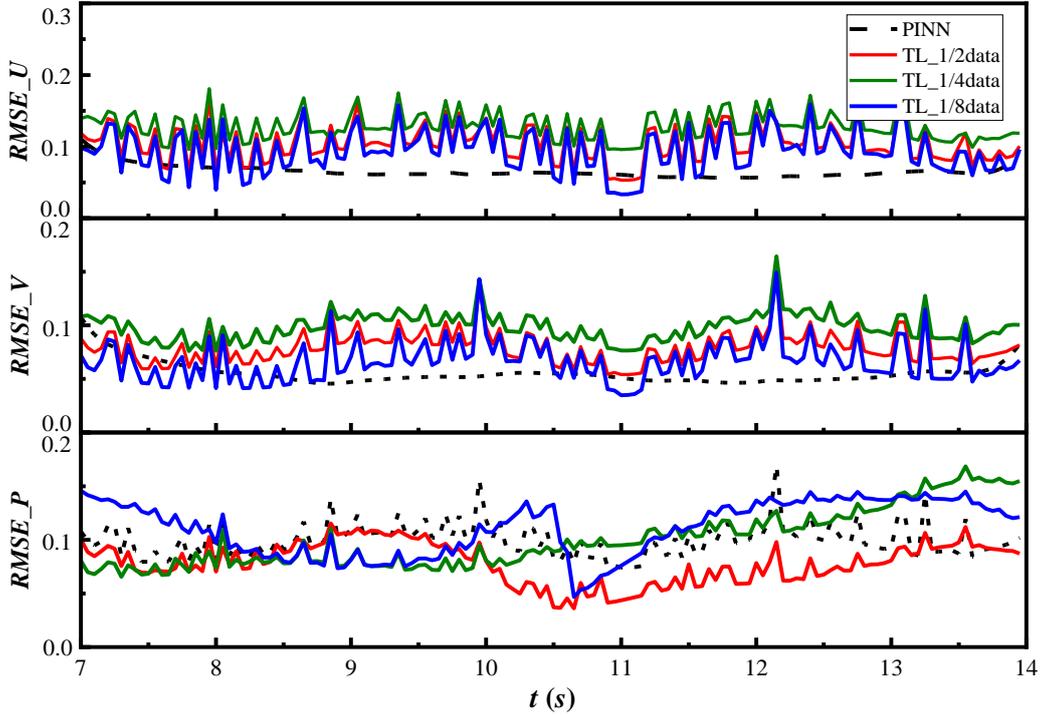

Fig. 13. The RMSE for the forecast of the velocity and pressure.

As Fig. 13 and Fig. 14 are illustrated, the amplitude of RMSE for the IL velocity $u$ of the four work conditions is 0.121, 0.163, 0.178 and 0.164 respectively. The amplitude of RMSE for the CF velocity $v$ of the four working conditions is 0.115, 0.124,

0.171 and 0.163 respectively. The amplitude of RMSE for the pressure *p* of the four work conditions is 0.178, 0.119, 0.173 and 0.146 respectively. And the amplitude of RMSE for the vibration displacement *r* of the four work conditions is 0.0007, 0.0013, 0.0021 and 0.0048 respectively. For the above working conditions, the RMSE for the forecast results of the velocity and pressure all fluctuate below 0.02, and the RMSE for the forecast results of the vibration displacement all fluctuate below 0.005, which satisfies our expected forecast accuracy.

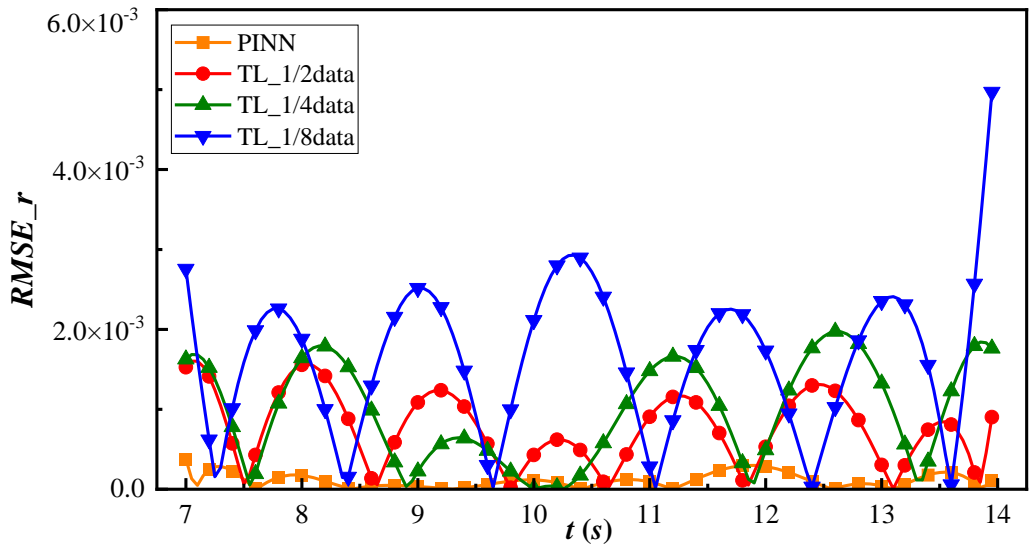

Fig. 14. The RMSE for the forecast of vibration displacement of the structure.

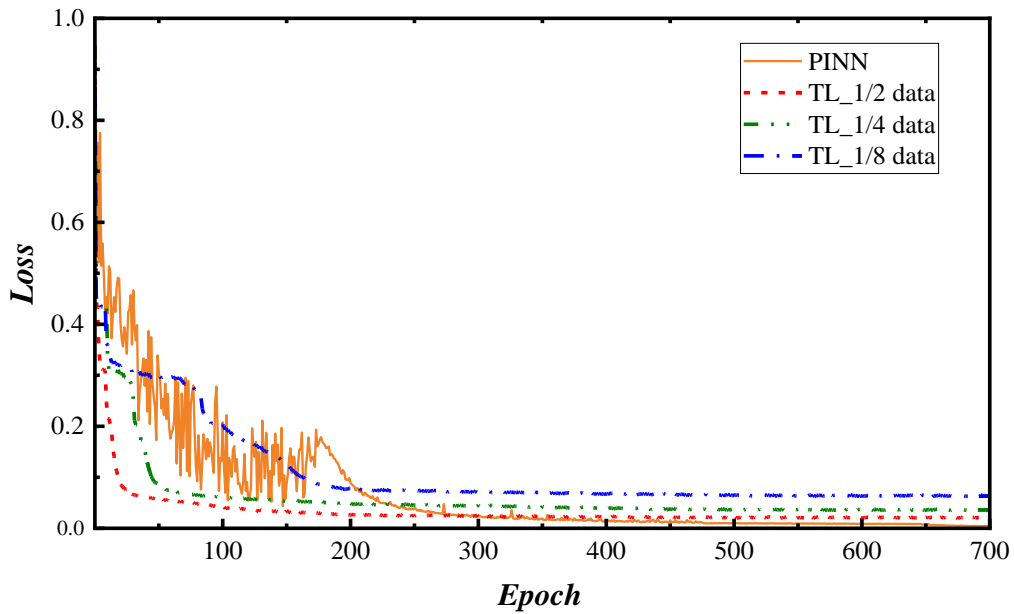

Fig. 15. The loss function of the four working conditions.

According to the forecast results, it is proved that the four working conditions all complete learning tasks with high accuracy in the target domain. In addition, the RMSE is easily affected by the position of monitoring points in different time periods, which cannot be ignored. This is why the error curves fluctuate constantly. Compared with the conventional PINN, the transfer learning model can receive high-precision forecast results for VIV with lower datasets cost, which proves that this proposed model has a significant data-saving ability. In addition, as is shown in Table 3, the computation time of working conditions 2-4 is approximately 1/2, 1/5 and 1/10 of that of working condition 1 respectively, which greatly decreases the time cost while ensuring forecast accuracy. At last, descent of the loss function in the process of model iteration is shown in Fig.15, the working condition 1-4 approximately tend to be stable at the 300th, 30th, 50th and 180th epoch respectively, it is clear to see that the loss function of work conditions 2-4 tend to converge faster with the help of the transfer learning.

In order to verify that the proposed model can also accurately forecast the coupling fluid forces on the structure, the forecast results of working condition 2-4 are taken into Eq. (5)- Eq. (6) to obtain the forecast fluid forces. Change in the BC of the fluid-structure interaction surface, as well as the distribution of the velocity and pressure are controlled at any time. To evaluate the forecast error more conveniently, Eq. (14)- Eq. (15) are used to convert the fluid forces into relevant coefficients, which are the lift coefficient $C_l$ and the drag coefficient $C_d$ respectively, as is shown in Fig. 16. Refer to other studies, the amplitude of lift coefficient $C_l$ is regarded as the comparison index, and the average of drag coefficient $C_d$ is regarded as the comparison index. In the numerical examples, the VIV occurs in the flow field with $Re$=100, thus the results of the above working conditions are compared with Braza et al. (2006) , which is illustrated in Table 4 explicitly.

$$C_l = \frac{2F_l}{\rho d U^2} \quad (14)$$

$$C_d = \frac{2F_d}{\rho d U^2} \quad (15)$$

There exist differences among the fluid force coefficient obtained by different experiments or numerical simulations, the amplitude of $C_l$ usually floats in the interval of [0.29, 0.41], and the average of $C_d$ usually floats in the interval of [1.22, 1.51]. As is shown in Table 4, the forecast results of the above working conditions all fall within the correct and reasonable range. It is proved that the transfer learning can forecast the coupling fluid force coefficients accurately with lower cost of datasets and time. In addition, the distribution of monitoring sites on the surface of the structure perhaps affects the accuracy of results, so the errors are all within acceptable limits.

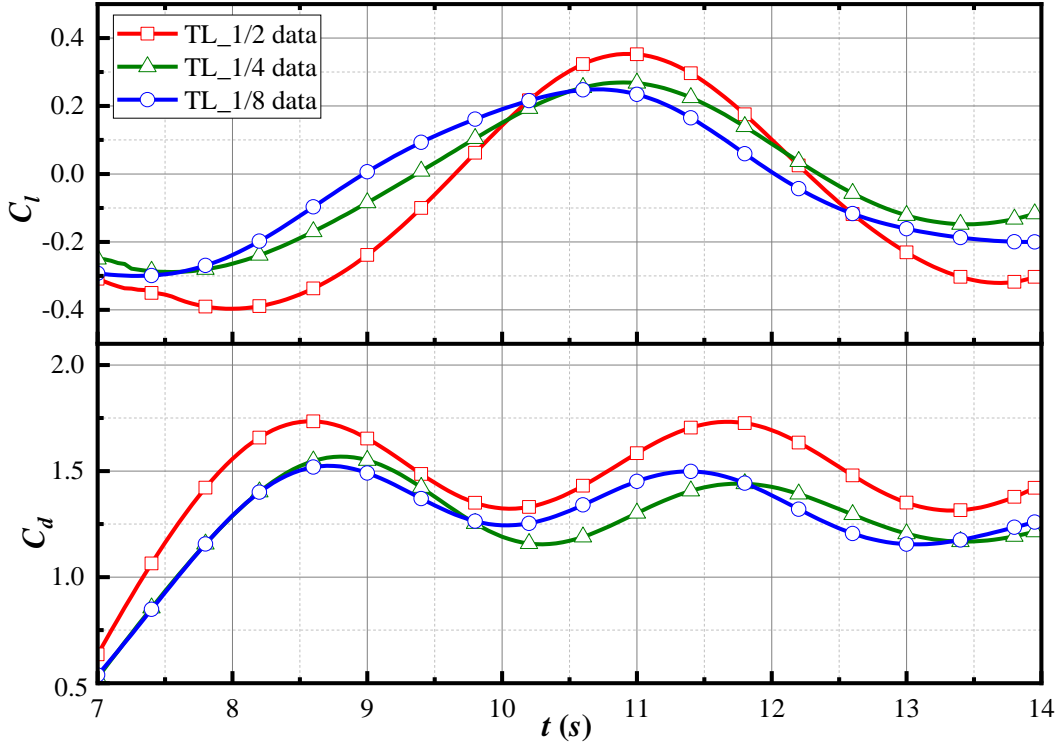

Fig. 16. The forecast results of the fluid force coefficient.

Table 4. Comparison of the lift coefficient $C_l$ and the drag coefficient $C_d$

| Serial number | Forecast $C_l$ | $C_l$ in the reference | Error of $C_l$ | Forecast $C_d$ | $C_d$ in the reference | Error of $C_d$ |
|---|---|---|---|---|---|---|
| Condition 2 | 0.398 | 0.355 | 0.043 | 1.463 | 1.372 | 0.091 |
| Condition 3 | 0.289 | 0.355 | 0.066 | 1.274 | 1.372 | 0.098 |
| Condition 4 | 0.301 | 0.355 | 0.054 | 1.279 | 1.372 | 0.093 |

## 6. Conclusions

In this paper, a transfer learning enhanced the physics-informed neural network model was proposed to break through the limit of computational efficiency for VIV analysis. By means of the model, the common characteristics knowledge shared by conventional PINN model for VIV is extracted from the source domain. Thereby it is feasible to perform diverse but related learning tasks more frugally in the target domain. Furthermore, 2D VIV experimental datasets collected by the PIV are used to verify the performance of the proposed method in forecasting the velocity and pressure of the flow field, as well as the vibration displacement of the structure and the coupling fluid forces. The main conclusions are obtained as follows:

(1) The four working conditions all can forecast the velocity and pressure of the flow field and the vibration displacement of the structure with high accuracy. The RMSE for the forecast results of the velocity and pressure all fluctuate below 0.02, and the RMSE for the forecast results of the vibration displacement all fluctuate below 0.005.

(2) The working conditions using transfer learning can also forecast the most interaction conditions of VIV correctly, which is the lift force $F_l$ and the drag force $F_d$. The fluid forces are transformed into the fluid force coefficients, and the forecast results of $C_l$ and $C_d$ both fall within the reasonable range.

(3) Compared with the conventional PINN, the transfer learning model for VIV demonstrates more substantial data-saving and time-saving ability. The Working conditions 2-4 using transfer learning only need 1/2，1/4 and 1/8 of the datasets in the target domain respectively. And the computation time of working conditions 2-4 is just 1/2, 1/5 and 1/10 of that of working condition 1. It is obvious that the transfer learning has a significant enhancement to PINN model for VIV to complete identical learning tasks.

## Acknowledgments

This research was supported by the Shanghai Municipal Science and Technology Major Project (2021SHZDZX0100).

## References

Bai, X., & Zhang, W. (2021). Machine Learning for Vortex Induced Vibration in Turbulent Flow. *arXiv preprint arXiv:2103.05818*.

Bau, D., Zhu, J. Y., Strobelt, H., Lapedriza, A., Zhou, B., & Torralba, A. (2020). Understanding the role of individual units in a deep neural network. *Proceedings of the National Academy of Sciences*, *117*(48), 30071-30078.

Bengio, Y. (2012, June). Deep learning of representations for unsupervised and transfer learning. In *Proceedings of ICML workshop on unsupervised and transfer learning* (pp. 17-36). JMLR Workshop and Conference Proceedings.

Braza, M., Chassaing, P. H. H. M., & Minh, H. H. (1986). Numerical study and physical analysis of the pressure and velocity fields in the near wake of a circular cylinder. *Journal of fluid mechanics*, *165*, 79-130.

Cai, S., Mao, Z., Wang, Z., Yin, M., & Karniadakis, G. E. (2021). Physics-informed neural networks (PINNs) for fluid mechanics: A review. *arXiv preprint*

*arXiv:2105.09506*.

Chen, J., Chen, J., Zhang, D., Sun, Y., & Nanehkaran, Y. A. (2020). Using deep transfer learning for image-based plant disease identification. *Computers and Electronics in Agriculture*, *173*, 105393.

Cheng, C., Meng, H., Li, Y. Z., & Zhang, G. T. (2021). Deep learning based on PINN for solving 2 DOF vortex induced vibration of cylinder. *Ocean Engineering*, *240*, 109932.

Dan, D., & Li, H. (2021). Monitoring, intelligent perception and early warning of vortex-induced vibration of suspension bridge.

Deng, L., Hinton, G., & Kingsbury, B. (2013, May). New types of deep neural network learning for speech recognition and related applications: An overview. In *2013 IEEE international conference on acoustics, speech and signal processing* (pp. 8599-8603). IEEE.

Geaur Rahman, M., & Zahidul Islam, M. (2021). A Framework for Supervised Heterogeneous Transfer Learning using Dynamic Distribution Adaptation and Manifold Regularization. *arXiv e-prints*, arXiv-2108.

Goswami, S., Anitescu, C., Chakraborty, S., & Rabczuk, T. (2020). Transfer learning enhanced physics informed neural network for phase-field modeling of fracture. *Theoretical and Applied Fracture Mechanics*, *106*, 102447.

Gupta, V., Choudhary, K., Tavazza, F., Campbell, C., Liao, W. K., Choudhary, A., & Agrawal, A. (2021). Cross-property deep transfer learning framework for enhanced predictive analytics on small materials data. *Nature communications*, *12*(1), 1-10.

Jin, X., Cai, S., Li, H., & Karniadakis, G. E. (2021). NSFnets (Navier-Stokes flow nets): Physics-informed neural networks for the incompressible Navier-Stokes equations. *Journal of Computational Physics*, *426*, 109951.

Kaur, T., & Gandhi, T. K. (2020). Deep convolutional neural networks with transfer learning for automated brain image classification. *Machine Vision and Applications*, *31*(3), 1-16.

Kim, G. Y., Lim, C., Kim, E. S., & Shin, S. C. (2021). Prediction of Dynamic Responses of Flow-Induced Vibration Using Deep Learning. *Applied Sciences*, *11*(15), 7163.

Li, X., Chen, H., Chen, B., Luo, X., Yang, B., & Zhu, Z. (2020). Investigation of flow pattern and hydraulic performance of a centrifugal pump impeller through the PIV method. *Renewable Energy*, *162*, 561-574.

Lim, D. H., & Kim, K. S. (2021). Development of Deep Learning-based Detection Technology for Vortex-Induced Vibration of a Ship's Propeller. *Journal of Sound and Vibration*, 116629.

Liu, W., Wang, Z., Liu, X., Zeng, N., Liu, Y., & Alsaadi, F. E. (2017). A survey of deep neural network architectures and their applications. *Neurocomputing*, *234*, 11-26.

Liu, X. Q., Jiang, Y., Liu, F. L., Liu, Z. W., Chang, Y. J., & Chen, G. M. (2021). Optimization Design of Fairings for VIV Suppression Based on Data-Driven Models and Genetic Algorithm. *China Ocean Engineering*, *35*(1), 153-158.

Liu, Z., Shen, J., Li, S., Chen, Z., Ou, Q., & Xin, D. (2021). Experimental study on high-mode vortex-induced vibration of stay cable and its aerodynamic countermeasures. *Journal of Fluids and Structures*, *100*, 103195.

Martini, S., Morgut, M., & Pigazzini, R. (2021). Numerical VIV analysis of a single elastically-mounted cylinder: Comparison between 2D and 3D URANS simulations. *Journal of Fluids and Structures*, *104*, 103303.

Neyshabur, B., Sedghi, H., & Zhang, C. (2020). What is being transferred in transfer learning?. *arXiv preprint arXiv:2008.11687*.

Nikoo, H. M., Bi, K., & Hao, H. (2018). Effectiveness of using pipe-in-pipe (PIP) concept to reduce vortex-induced vibrations (VIV): Three-dimensional two-way FSI analysis. *Ocean Engineering*, *148*, 263-276.

Pan, S. J., & Yang, Q. (2009). A survey on transfer learning. *IEEE Transactions on knowledge and data engineering*, *22*(10), 1345-1359.

Raynaud, G., Houde, S., & Gosselin, F. P. (2021). ModalPINN: an extension of Physics-Informed Neural Networks with enforced truncated Fourier decomposition for periodic flow reconstruction using a limited number of imperfect sensors. *arXiv preprint arXiv:2108.08929*.

Raissi, M., Perdikaris, P., & Karniadakis, G. E. (2019). Physics-informed neural networks: A deep learning framework for solving forward and inverse problems involving nonlinear partial differential equations. *Journal of Computational Physics*, *378*, 686-707.

Raissi, M., Wang, Z., Triantafyllou, M. S., & Karniadakis, G. E. (2019). Deep learning of vortex-induced vibrations. *Journal of Fluid Mechanics*, *861*, 119-137.

Shu, Y., & Li, B. (2021). Surface Defect Detection and Recognition Method for Multi-Scale Commutator Based on Deep Transfer Learning. *Arabian Journal for*

*Science and Engineering*, 1-12.

Sun, L., Gao, H., Pan, S., & Wang, J. X. (2020). Surrogate modeling for fluid flows based on physics-constrained deep learning without simulation data. *Computer Methods in Applied Mechanics and Engineering*, *361*, 112732.

Wang, W., Song, B., Mao, Z., Tian, W., Zhang, T., & Han, P. (2020). Numerical investigation on vortex-induced vibration of bluff bodies with different rear edges. *Ocean Engineering*, *197*, 106871.

Williamson, C. H. K., & Govardhan, R. (2008). A brief review of recent results in vortex-induced vibrations. *Journal of Wind engineering and industrial Aerodynamics*, *96*(6-7), 713-735.

Wong, E. W. C. (2018). A simplified method to predict fatigue damage of TTR subjected to short-term VIV using artificial neural network. *Advances in engineering software*, *126*, 100-109.

Wu, W., Sun, H., Lv, B., & Bernitsas, M. M. (2018). Modelling of a hydrokinetic energy converter for flow-induced vibration based on experimental data. *Ocean engineering*, *155*, 392-410.

Wu, Z., Jiang, H., Zhao, K., & Li, X. (2020). An adaptive deep transfer learning method for bearing fault diagnosis. *Measurement*, *151*, 107227.

Ye, R., & Dai, Q. (2021). Implementing transfer learning across different datasets for time series forecasting. *Pattern Recognition*, *109*, 107617.

Zafar, T., & Wang, Z. (2020). Time-dependent reliability prediction using transfer learning. *Structural and Multidisciplinary Optimization*, *62*(1), 147-158.

Zhang, W., Du, Y., Yoshida, T., & Yang, Y. (2019). DeepRec: A deep neural network approach to recommendation with item embedding and weighted loss function. *Information Sciences*, *470*, 121-140.

Zhuang, F., Qi, Z., Duan, K., Xi, D., Zhu, Y., Zhu, H., & He, Q. (2020). A comprehensive survey on transfer learning. *Proceedings of the IEEE*, *109*(1), 43-76. 5